\documentstyle[12pt]{article}
\newcommand{\hs}{\hspace{.15cm}}
\def\mm#1{ m_{\tilde #1}^2  }
\def\smm#1{   m_{ #1}^2   }
\def\s#1{   \theta_{ #1}  }
\def\ss#1#2{\theta_{{ #1}{#2}}    }
\def\t#1{   \theta_{\tilde #1}  }
\def\st#1#2{   \theta_{{ #1}{\tilde #2}}   }
\def\tt#1#2{   \theta_{{\tilde #1}{\tilde #2}}   }
\def\tt#1#2{   \theta_{{\tilde #1}{\tilde #2}}   }

\def\ttt#1#2#3{   \theta_{{\tilde #1}{\tilde #2}{\tilde #3}}   }
\begin{document}
\begin{titlepage}
\begin{centering}
{\large{\bf Low
Energy Thresholds and the Renormalization Group
in the MSSM}}\\
\vspace{.5in}
{{\bf A. B. Lahanas } \hspace{.1cm}}$^{\dag}$ \\
\vspace{.2in}%
University of Athens, Physics Department,
%\vspace{.05in}
Nuclear
and Particle Physics Section\\
\vspace{.05in}
Ilissia, GR - 15771  Athens,
GREECE\\
\vspace{.2in}
and \\
\vspace{.2in}
{{\bf K. Tamvakis} \hspace{.1cm}}$^{\ddag}$  \\
\vspace{.2in}
Division of Theoretical Physics,
University of Ioannina
\\
Ioannina, GR - 451 10\,GREECE\\
\vspace{.4in}
{\bf Abstract}\\
\vspace{.1in}
\end{centering}
{\noindent
We derive the 1-loop Renormalization Group Equations for the
parameters
of the Minimal Supersymmetric Standard Model (MSSM) taking into
account
the successive decoupling of each sparticle below its threshold.
This
is realized by a step function at the level of each graph
contributing
to the Renormalization Group Equations.}

{\vspace*{1.2in}}
{\hspace{.5cm}}{\it  Dedicated to the memory of Spyros Vlassopulos ,
colleague and friend}
\\
\vspace*{.5in}
%
%\paragraph{}%

\par
\vspace{4mm}
\begin{flushleft}
IOA-314/94\\
UA/NPPS -
16/1994\\
%
%hep-ph/9412XXX \\
\end{flushleft}
\rule[0.in]{4.5in}{.01in}%
\\
E-mail:$^{{\hspace{.2cm}}\dag}$ { \hspace{.1cm}
alahanas@atlas.uoa.ariadne-t.gr} ,\quad
$^{{\hspace{.2cm}}\ddag}$ {
\hspace{.1cm}
tamvakis@cc.uoi.gr}
\end{titlepage}

%%%%%%%%%%%%%%%%%%%%%%%%%%%%%%%%%%%%%%%%%%%%%%%%
\par
The softly broken version of the Minimal Supersymmetric Standard Model
{\mbox{(MSSM)$^{\cite{Nilles}}$ }
is well known to lead to Electroweak Symmetry breaking through radiative
corrections$^{\cite{Iban}}$~.
The study of radiative corrections is most conveniently
done by the use of the Renormalization Group Equations (RGE's) for the running
parameters of the model. Ultimately these energy dependent parameters should
be related to the physical ones. One popular way to realize this program
$^{\cite{Ross}}$
exclusively in the framework of the (RGE's) is the following:

We derive the Renormalization Group Equations for running masses
and couplings in the ${\overline{DR}}$ scheme .As we come down in energy
and encounter the heaviest particle threshold we switch to another "effective"
theory$^{\cite{Weinberg}}$
from which the heavy particle is removed and so on. Although  the
 ${\overline{DR}}$  scheme is mass independent , we enforce the Decoupling
Theorem$^{\cite{Appel}}$
at the level of the (RGE's) by replacing
the full theory by a succession of effective theories. For a particular
running mass $m(Q)$ this treatment of its  Renormalization Group Equation
will stop when we encounter the corresponding physical mass determined by the
condition $m(m_{ph})=m_{ph}$. Both the step function approximation on the
RGE's and this last condition for the physical
mass are approximations that keep the leading logarithmic part in 2-point
functions and ignore the constant part. The popularity of this scheme is
based on the fact that one stays only within the Renormalization Group
Equations$^{\cite{Falck}}$
and does not have to consider the finite parts of the Green
functions.

In the present paper we derive in the ${\overline{DR}}$ scheme
the 1-loop RGE's for both dimensionless (gauge and
Yukawa couplings) and dimensionful parameters (soft masses and cubic
couplings). At the level of each graph we enforce decoupling by inserting a
theta function ${{\theta}_m} \equiv \theta(Q^2-m^2)$ that counts the
contribution of a
particle of mass $m$ at energies $Q>m$ . In order for such a program to be
carried out we should calculate the infinities of the two and three point
functions of all particles involved. The use of the superfield formalism
in a non-supersymmetric theory, although possible through the utilization
of spurious superfields , becomes difficult due to the successive decoupling
occurring at the component level. One has to resort to the component
formulation of the theory  where the non-Renormalization Theorem does not
apply. The details of the calculation along with the predictions of the
MSSM for the sparticle mass spectrum using the modified RGE's described
above will be presented elsewhere$^{\cite{Lahanas}}$.

The parameters of the MSSM
are defined by the superpotential (suppressing all indices)
\begin{equation}
{\cal W}=Y_u {Q}{H}_2 {U}^c
       +Y_d {Q} {H}_1 {D}^c
       +Y_e {L} {H}_1 {E}^c
        +\mu  {H}_1 {H}_2
%\quad\quad \epsilon_{12}=+1,
\end{equation}
%%%%%
and the soft supersymmetry-breaking interaction Lagrangian
%%%%%%%%
\begin{eqnarray}
- {\cal L}_{{soft}}&=&{\sum_{i}} m_i^2 |\Phi_i|^2
         + (Y_u A_u {\tilde Q} H_2 {\tilde U}^c
         + Y_d A_d {\tilde Q}  H_1 {\tilde D}^c
         + Y_e A_e {\tilde L} H_1 {\tilde E}^c+h.c.)\nonumber\\
                         &+&(  \mu B H_1 H_2 +h.c.)
         +\frac {1}{2} \sum_a M_a {\bar{\lambda}}_a \lambda_a .
\end{eqnarray}
%%%%%%%%
The beta functions for the gauge and Yukawa couplings are,
with $t \equiv \log(\frac {Q^2}{M_{GUT}^2})$ and keeping only the Yukawa
couplings $Y_{t,b,\tau}$ of the third generation fermions ,
\newpage
%%%%%%%%%%%%
%$$
\begin{eqnarray}
{\frac {dg_i}{dt}}&\equiv&{\beta(g_i)} ={\frac {b_i}{ {2{(4\pi)}^2 }}  }
T_i\,{g_i}^3 \quad , \quad i=1,2,3 \\
b_{1,2,3}& = & {\frac {33}{5}} , 1, -3 \nonumber \\
{\frac {dY_\tau}{dt}}& \equiv& {\beta(Y_\tau)}={\frac {Y_\tau}{{(4\pi)}^2} }
  \{- {3 \over 2}T_{\tau2} {g_2}^2 - {\frac {9}{10}}T_{\tau1} {g_1}^2
  +2 T_{\tau\tau} {Y_\tau}^2 + {3 \over 2} {Y_b}^2  \}  \\
{\frac {dY_b}{dt}}& \equiv& {\beta(Y_b)}=  \nonumber \\
& &{\frac {Y_b}{{(4\pi)}^2} }
  \{-{8\over3}T_{b3} {g_3}^2 - {3 \over 2} T_{b2} {g_2}^2
  -{\frac {7}{30}}T_{b1} {g_1}^2
+{1 \over 2} T_{bt} {Y_t}^2 + 3T_{bb} {Y_b}^2+ {1 \over 2} {Y_\tau}^2 \}  \\
{\frac {dY_t}{dt}}& \equiv& {\beta(Y_t)}={\frac {Y_t}{{(4\pi)}^2} }
  \{-{8\over3}T_{t3} {g_3}^2-{3 \over 2}  T_{t2} {g_2}^2
  - {\frac {13}{30}}T_{t1} {g_1}^2
  + 3T_{tt} {Y_t}^2 + {1 \over 2}T_{tb} {Y_b}^2  \}
\end{eqnarray}
%%%%%%%%%%%%%%

The threshold coefficients $T_i,\, T_{ei}, etc$ appearing in the expressions
above  are shown in table I.

The beta functions for the cubic couplings are{\footnote{\hspace{.1cm}
$\tilde{G},\tilde{W},\tilde{B}$
denote the $SU(3), SU(2) $ and $U(1)$ gauge fermions respectively. \\
\hspace*{.8cm} $\tilde{H_1},\tilde{H_2}$ are  Higgs fermions ( Higgsinos ).}}
%\newpage
%%%%%%%%%%%%
\begin{eqnarray}
 {\frac {dA_\tau}{dt} }&=&{\frac {1}{{(4\pi)}^2} }   \{
-3{g_2}^2 {M_2} \tt{W}{H_1} -{3 \over 5}{g_1}^2 {M_1}
{(2+\t{H_1})\t{B}}  \nonumber  \\  \nonumber  \\
&+&3{Y_b}^2 {A_b} \tt{D}{Q}+4{Y_\tau}^2 {A_\tau}+{A_\tau}[ {Z_{\tau 1}}{g_1}^2
+ {Z_{\tau 2}}{g_2}^2 +{Z_{\tau \tau}}{Y_\tau}^2]    \}   \\  \nonumber \\
 {\frac {dA_b}{dt} }&=&{\frac {1}{{(4\pi)}^2} }   \{
-{16 \over 3}{g_3}^2 {M_3} \t{G} -3{g_2}^2 {M_2} \tt{W}{H_1}
-{1 \over 30}{g_1}^2 {M_1} {(-4+18 \t{H_1}) \t{B} }  \nonumber \\ \nonumber \\
&+&{Y_\tau}^2 {A_\tau} \tt{E}{L}
+ {Y_t}^2 {A_t} \st{H_2}{U}
+6{Y_b}^2 {A_b}        \nonumber  \\  \nonumber  \\
&+& {A_b}[{Z_{b3}} {g_3}^2+{Z_{b2}} {g_2}^2+{Z_{b1}} {g_1}^2
+ {Z_{bt}} {Y_t}^2   +   {Z_{bb}} {Y_b}^2  ]    \}     \\ \nonumber  \\
 {\frac {dA_t}{dt} }&=&{\frac {1}{{(4\pi)}^2} }   \{
-{16 \over 3}{g_3}^2 {M_3} \t{G} -3{g_2}^2 {M_2} \tt{W}{H_2}
-{1 \over 15}{g_1}^2 {M_1} {(4+9 \t{H_2}) \t{B} }  \nonumber \\ \nonumber \\
&+& 6{Y_t}^2 {A_t} \st{H_1}{U}
+{Y_b}^2 {A_b}\st{H_1}{D}         \nonumber  \\  \nonumber  \\
&+& {A_t}[{Z_{t3}} {g_3}^2+{Z_{t2}} {g_2}^2+{Z_{t1}} {g_1}^2
+ {Z_{tt}} {Y_t}^2   +   {Z_{tb}} {Y_b}^2  ]    \}     \\  \nonumber
\end{eqnarray}
%%%%%%%%%%%%
In our notation \, $\theta_{ab} \equiv \theta_{a}  \theta_{b} $ \,
$\theta_{abc} \equiv \theta_{a}  \theta_{b} \theta_{c}    $   . \@
In all the expressions throughout this paper we assume that the Yukawa
couplings are diagonal in family space. The coefficients $Z_{qi}$ in the
expressions for the cubic couplings above  vanish above all thresholds
and in this case one recovers the well known RGEs .

The beta functions for the  scalar masses are given by the following
RGEs {\footnote{The RGEs for the soft masses of the squarks and sleptons
presented here refer to the third generation of fermions.
For the first two generations the Yukawa couplings should be set to zero}}  .
The masses $m_{1,2}^2$ appearing in the RGEs below refer to the
Higgs masses squared which are related to the their soft masses
$m_{H_1,H_2}^2$
and the mixing parameter $\mu$ by  $m_{1,2}^2=m_{H_1,H_2}^2+{\mu}^2 $.
%%%%%%%%%%%%
\begin{eqnarray}
{\frac {d{\mm{Q}}} {dt}  }      % & \equiv& {\beta(m_{\mm{Q}}) }
&=&  {\frac {1}{{(4\pi)}^2} }  \{- [{8\over3}{g_3}^2(\t{Q}-\t{G})
+{3\over2}{g_2}^2(\t{Q}-\t{W})
+{1 \over 30}{g_1}^2(\t{Q}-\t{B}) ] \, {\mm{Q}}
\nonumber \\  \nonumber  \\
&-&{16\over3}{g_3}^2 {M_3^2} \t{G}-3{g_2}^2 {M_2^2} \t{W}
-{1 \over 15}{g_1}^2 {M_1^2} \t{B} + {1 \over 10}{{g_1}^2} S
\nonumber \\ \nonumber \\
&+&{Y_t}^2 [\mm{Q} \t{H_2} +\mm{U} \t{U}+\smm{2} \s{H_2}+A_t^2 \st{H_2}{U}
        + {\mu^2} (\st{H_1}{U}-2\t{H_2})]   \}   \nonumber \\ \nonumber \\
&+&{Y_b}^2 [\mm{Q} \t{H_1} +\mm{D} \t{D}+\smm{1} \s{H_1}+A_b^2 \st{H_1}{D}
        + {\mu^2} (\st{H_2}{D}-2\t{H_1}) ]   \}  \\ \nonumber  \\
{\frac {d{\mm{U}}} {dt}  }      % & \equiv& {\beta(m_{\mm{Q}}) }
&=&  {\frac {1}{{(4\pi)}^2} }  \{-[{8\over3}{g_3}^2(\t{U}-\t{G})
%+{3\over2}{g_2}^2(\t{Q}-\t{W})
+{8 \over 15}{g_1}^2(\t{U}-\t{B}) ] \, {\mm{U}}
\nonumber \\  \nonumber  \\
&-&{16\over3}{g_3}^2 {M_3^2} \t{G}%-3{g_2}^2 {M_2^2} \t{W}
-{16 \over 15}{g_1}^2 {M_1^2} \t{B} - {2 \over 5}{{g_1}^2} S
\nonumber \\ \nonumber \\
&+&2{Y_t}^2 [\mm{U} \t{H_2} +\mm{Q} \t{Q}+\smm{2} \s{H_2}+A_t^2 \st{H_2}{Q}
        + {\mu^2} (\st{H_1}{Q}-2\t{H_2})]   \}   \\ \nonumber \\
{\frac {d{\mm{D}}} {dt}  }      % & \equiv& {\beta(m_{\mm{Q}}) }
&=&  {\frac {1}{{(4\pi)}^2} }  \{- [{8\over3}{g_3}^2(\t{D}-\t{G})
%+{3\over2}{g_2}^2(\t{D}-\t{W})
+{2 \over 15}{g_1}^2(\t{D}-\t{B}) ] \, {\mm{D}}
\nonumber \\  \nonumber  \\
&-&{16\over3}{g_3}^2 {M_3^2} \t{G}%+3{g_2}^2 {M_2^2} \t{W}
-{4 \over 15}{g_1}^2 {M_1^2} \t{B} + {1 \over 5}{{g_1}^2} S
\nonumber \\ \nonumber \\
&+&2{Y_b}^2 [\mm{D} \t{H_1} +\mm{Q} \t{Q}+\smm{1} \s{H_1}+A_b^2 \st{H_1}{Q}
        + {\mu^2} (\st{H_2}{Q}-2\t{H_1})]   \}   \\ \nonumber \\
{\frac {d{\mm{L}}} {dt}  }      % & \equiv& {\beta(m_{\mm{Q}}) }
&=&  {\frac {1}{{(4\pi)}^2} }  \{- [
{3\over2}{g_2}^2(\t{L}-\t{W})
+{3 \over 10}{g_1}^2(\t{L}-\t{B}) ] \, {\mm{L}}
\nonumber \\  \nonumber  \\
&-&3{g_2}^2 {M_2^2} \t{W}
-{3 \over 5}{g_1}^2 {M_1^2} \t{B} - {3 \over 10}{{g_1}^2} S
\nonumber \\ \nonumber \\
&+&{Y_\tau}^2 [\mm{L} \t{H_1} +\mm{E} \t{E}+\smm{1} \s{H_1}
+A_\tau^2  \st{H_1}{E}
        + {\mu^2} (\st{H_2}{E}-2\t{H_1})]   \}   \\ \nonumber \\
{\frac {d{\mm{E}}} {dt}  }      % & \equiv& {\beta(m_{\mm{Q}}) }
&=&  {\frac {1}{{(4\pi)}^2} }  \{- [
{6 \over 5}{g_1}^2(\t{E}-\t{B}) ] \, {\mm{E}}
\nonumber \\  \nonumber  \\
&-&{12 \over 5}{g_1}^2 {M_1^2} \t{B} + {3 \over 5}{{g_1}^2} S
\nonumber \\ \nonumber \\
&+&2{Y_\tau}^2 [\mm{E} \t{H_1} +\mm{L} \t{L}+\smm{1} \s{H_1}
+A_\tau^2  \st{H_1}{L}
        + {\mu^2} (\st{H_2}{L}-2\t{H_1})]   \}   \\ \nonumber \\
{\frac {d{\smm{1}}} {dt}  }      % & \equiv& {\beta(m_{\mm{Q}}) }
&=&  {\frac {1}{{(4\pi)}^2} }  \{- [
{3\over2}{g_2}^2(\s{H_1}-\tt{H_1}{W})
+{3 \over 10}{g_1}^2(\s{H_1}-\tt{H_1}{B} ) ] \, {\smm{1}}
\nonumber \\  \nonumber  \\
&-&3{g_2}^2( {M_2^2+\mu^2}) \tt{H_1}{W}
-{3 \over 5}{g_1}^2( {M_1^2+\mu^2}) \tt{H_1}{B}-{3 \over 10}{{g_1}^2} S
\nonumber \\ \nonumber \\
&+&{Y_\tau}^2 [\smm{1} +\mm{L} \t{L}+\mm{E} \t{E}+A_\tau^2  \tt{L}{E} ]
\nonumber \\ \nonumber \\
&+&3{Y_b}^2 [\smm{1} +\mm{Q} \t{Q}+\mm{D} \t{D}+A_b^2  \tt{Q}{D} ]
+ 3{{Y_t}^2} {\mu^2} \tt{Q}{U}
   \}   \\ \nonumber \\
{\frac {d{\smm{2}}} {dt}  }      % & \equiv& {\beta(m_{\mm{Q}}) }
&=&  {\frac {1}{{(4\pi)}^2} }  \{- [
{3\over2}{g_2}^2(\s{H_2}-\tt{H_2}{W})
+{3 \over 10}{g_1}^2(\s{H_2}-\tt{H_2}{B} ) ] \, {\smm{2}}
\nonumber \\  \nonumber  \\
&-&3{g_2}^2( {M_2^2+\mu^2}) \tt{H_2}{W}
-{3 \over 5}{g_1}^2( {M_1^2+\mu^2}) \tt{H_2}{B}+{3 \over 10}{{g_1}^2} S
\nonumber \\ \nonumber \\
&+&3{Y_t}^2 [\smm{2} +\mm{Q} \t{Q}+\mm{U} \t{U}+A_t^2  \tt{Q}{U} ]
\nonumber \\ \nonumber \\
&+&3{Y_b}^2 {\mu^2} \tt{Q}{D}
+ {{Y_\tau}^2} {\mu^2} \tt{E}{L}       \}
 \\ \nonumber
\end{eqnarray}
%%%%%%%%%%%%%%

The quantity $S$ appearing in the equations above is defined as
%%%%%%%%%%%%%%%%%%%
\begin{equation}
S \equiv \hs   Tr \hs  \{ {Y\over 2} \hs  \theta_m \hs  m^2 \}
% S \equiv \hs  {1 \over 2} \hs {\sum_{i=1}} \hs {Y_i} \hs
%{\theta} ({\mu}^2-{m^2_i}(\mu)) \hs  { m^2_i}(\mu)
\end{equation}
%%%%%%%%%%%%%%%%%%%
This vanishes if  universal boundary conditions are assumed for all soft
scalar masses involved at the unification scale, as long as we are above
all particle thresholds. This is due to the fact that
$S$ is multiplicatively renormalized.

The RGEs given above refer to the third generation .
For the first two generations we have just to make the appropriate
replacements for the Yukawa couplings  which, due to their smallness, we have
assumed zero.

 For the Higgs and Higgsino mixing parameters $m_3^2 \equiv B \mu$
and $\mu$ respectively we {\mbox{have,}}
%%%%%%%%%%%%
\begin{eqnarray}
{\frac {d{\smm{3}}} {dt}  }      % & \equiv& {\beta(m_{\mm{Q}}) }
&=&  {\frac {1}{{(4\pi)}^2} }  \{ [
-{3 \over 4}{g_2}^2
{ (\s{H_1}+\s{H_2}+2\ss{H_1}{H_2}-\tt{H_1}{W}-\tt{H_2}{W})}
\nonumber \\  \nonumber  \\
&-&{3 \over 20}{g_1}^2
 { (\s{H_1}+\s{H_2}+2\ss{H_1}{H_2}-\tt{H_1}{B}-\tt{H_2}{B})}
\nonumber \\  \nonumber  \\
&+&{3 \over 2}{Y_t}^2+{3 \over 2}{Y_b}^2 +{1 \over 2}{Y_\tau}^2 ]
\,\,\, {\smm{3}}
\nonumber \\  \nonumber  \\
&+& \mu \, [- 3{g_2}^2 {M_2} \ttt{H_1}{H_2}{W}
-{3 \over 5}{g_1}^2{M_1} \ttt{H_1}{H_2}{B}
\nonumber \\  \nonumber  \\
&+&3 A_t{Y_t}^2 \tt{Q}{U}+3 A_b{Y_b}^2 \tt{Q}{D}
+A_\tau {Y_\tau}^2 \tt{L}{E} ]  \}
\end{eqnarray}
%\newpage
\begin{eqnarray}
{\frac {d{\mu}} {dt}  }      % & \equiv& {\beta(m_{\mm{Q}}) }
&=&  {\frac {1}{{(4\pi)}^2} }  \{
{3 \over 8}{g_2}^2 (\t{H_1}+\t{H_2}-8\tt{H_1}{H_2}+\st{H_1}{W}+\st{H_2}{W})
\nonumber \\  \nonumber  \\
&+&{3 \over 40}{g_1}^2(\t{H_1}+\t{H_2}-8\tt{H_1}{H_2}+\st{H_1}{B}+\st{H_2}{B})
\nonumber \\  \nonumber  \\
&+&{3 \over 4}{Y_b}^2(\t{Q}+\t{D})+{3 \over 4}{Y_t}^2(\t{Q}+\t{U})
+{1 \over 4}{Y_\tau}^2(\t{L}+\t{E}) \} \, \mu
\end{eqnarray}
%%%%%%%%%%%%%%
Finally the beta functions for the three gaugino masses are
%%%%%%%%%%%%%%%%%%%%%%%%%%%%%
\begin{eqnarray}
{\frac {d{M_i}} {dt}  }      % & \equiv& {\beta(m_{\mm{Q}}) }
={S_i}\hs {\frac {b_i}{{(4\pi)}^2} } \hs {g_i^2} \hs { M_i} \hs , \hs i=1,2,3
\end{eqnarray}
%%%%%%%%%%%%%%%%%%%%%%%%%%%%%
where $b_i$ are the beta function coefficients of the gauge couplings above
all particle thresholds , i.e
\begin{eqnarray}
b_{1,2,3}={\frac {33}{5}},1,-3
\end{eqnarray}
%%%%%%%%%%%%%%%%%%%%%%%%%%%%%%%%%%%
and $S_i$ are threshold function coefficients given by,
\begin{eqnarray}
S_3&=&-3 \hs  \t{G}
-{1 \over 6} \hs{\sum_{i=1}^{N_g}} \hs (2\t{Q_i}+\t{U_i}+\t{D_i})
\\ \nonumber \\
S_2&=& -6 \hs \t{W}-{1 \over 2} \hs {\sum_{i=1}^{N_g}} \hs (3\t{Q_i}+\t{L_i})
-{1 \over 2}(\st{H_1}{H_1}+\st{H_2}{H_2})
\\ \nonumber \\
S_1&=&{1 \over 11}\hs [\hs {\sum_{i=1}^{N_g}}  \hs
({1 \over 6}\t{Q_i}+{4 \over 3}\t{U_i} +
{1 \over 3}\t{D_i}+{1 \over 2}\t{L_i}+\t{E_i})
+{1 \over 2}(\st{H_1}{H_1}+\st{H_2}{H_2})  ]
\end{eqnarray}
%%%%%%%%%%%%%%%%%%%%%%%%%%%%%
Electroweak symmetry breaking effects , among which especially those
expressed directly through $m_t$ might play a role, have not been included
above . Note that the threshold effects computed above at the one loop
level are expected to be of the same order of magnitude as the standard
2-loop contributions to the RGE's$^{\cite{Lahanas}}$. Work on these subjects
is in progress.
%\par
%\vspace*{1in}
\newpage
\noindent
{\bf Acknoweledgments}\\ \\
We both thank the CERN Theory Division for hospitality during a short visit
in which part of this work was completed. K.T. also acknowledges
illuminating conversations with C. Savoy and I. Antoniadis during a visit
at Saclay in the framework of the EEC Human Capital and Mobility Network
"Flavourdynamics" (CHRX-CT93-0132). A.B.L.  acknoweledges support by the
EEC Science Program SC1-CT92-0792.
%\newpage

%%%%%%%%%%%%%%%%%%%%%%%%%%%%%%%%%%%%%%%%%%%%%%%%%

\newpage
{\bf Table Captions}

\vspace{1.cm}
{\bf Table I}:\quad Threshold coefficients appearing in the renormalization
group equations \\
of the gauge and Yukawa couplings. Above all thresholds these become
equal to unity.

\vspace{1.cm}
{\bf Table II}:\quad Threshold coefficients appearing in the renormalization
group equations \\
of the trilinear scalara couplings. Above all thresholds these are
vanishing.

%%%%%%%%%%%%%%%%%%%%%%
\begin{center}
\begin{tabular}{|l|}\hline
\multicolumn{1}{|c|}{\bf TABLE I} \\ \hline \\
$T_1={1 \over 33} [ 20 +
 {{\theta_{{\tilde H}_1}}+{\theta_{{\tilde H}_2}} } +
 {1 \over 2} (  {{\theta_{{H}_1}}+{\theta_{{ H}_2}} } )+
 {\sum_{i=1}^{3}}( {1 \over 2}
 {\theta_{{\tilde L}_i}}+ {\theta_{{\tilde E}_i}}+
 {1 \over 6}{\theta_{{\tilde Q}_i}}+{4 \over 3}{\theta_{{\tilde U}_i}}+
{1 \over 3}{\theta_{{\tilde D}_i}}          ) ]  $
\\  \\
$T_2=-{10 \over 3}+{4 \over 3}{\theta_{{\tilde W}}} +
{ 1\over 3}({{\theta_{{\tilde H}_1}}+{\theta_{{\tilde H}_2}} }        )+
{1 \over 6} (  {{\theta_{{H}_1}}+{\theta_{{ H}_2}} } ) +
 {1 \over 6}{\sum_{i=1}^{3}}(
3{\theta_{{\tilde Q}_i}}+{\theta_{{\tilde L}_i}}) $ \\   \\
$T_3={7 \over 3}-{2 \over 3}{\theta_{{\tilde G}}}-{1 \over 18}
{\sum_{i=1}^{3}}(2{\theta_{{\tilde Q}_i}}+{\theta_{{\tilde D}_i}}+
{\theta_{{\tilde U}_i}})$\\  \\  \hline \\

${T_{\tau 2}}={1 \over 4}[ -1+4{\theta_{{ H}_1}}-2{\theta_{ {{\tilde H}_1}
{\tilde W}}}-{\theta_{  {\tilde L}{\tilde W} } } +
4 {\theta_{   {{\tilde H}_1} {\tilde L}{\tilde W}  }   }  ] $ \\  \\

${T_{\tau 1}}={1 \over 12}[ 11-4{\theta_{ {\tilde B}{\tilde E} }}
+8 {\theta_{ {\tilde B} {\tilde E}{{\tilde H}_1} } }
-2 {\theta_{   {\tilde B}{{\tilde H}_1} } }+4{\theta_{{ H}_1}}
-{\theta_{  {\tilde B} {\tilde L}   }}
-4{\theta_{  {\tilde B} {\tilde L}{{\tilde H}_1}  }  } ] $ \\ \\

${T_{\tau \tau}}={1 \over 8}[ 2+\tt{H_1}{E}
+3{\theta_{{ H}_1}}+2{\theta_{  {\tilde L}{{\tilde H}_1}  }   }
   ] $ \\  \\  \hline  \\

${T_{b3}}={1 \over 4}[6-{\theta_{{\tilde G}{\tilde D}}}
-{\theta_{   {\tilde G}{\tilde Q} } }]$ \\ \\

${T_{b 2}}={1 \over 4}[ -1+4{\theta_{{ H}_1}}-2{\theta_{   {{\tilde H}_1}
{\tilde W} }}  -{\theta_{  { \tilde Q} {\tilde W}  } } +
4 {\theta_{  {{\tilde H}_1}{\tilde Q}{\tilde W}  } }] $ \\  \\

${T_{b 1}}={1 \over 28}[-21-4{\theta_{   {\tilde B}{\tilde D}   }}
-18{\theta_{   {\tilde B}{{\tilde H}_1}  }  }
+24{\theta_{ {{\tilde H}_1}{\tilde D}{\tilde B}     }}
+36{\theta_{{H}_1}}-{\theta_{   {\tilde B}{\tilde Q}   }}
+12{\theta_{  {{\tilde H}_1}{\tilde Q}{\tilde B}  } }]$ \\ \\

${T_{bt}}={1 \over 2}[\s{H_2}+\tt{ H_2}{ U}] $  \\ \\

${T_{bb}}={1 \over 12}[6+\tt{D}{H_1}+3\s{H_1}+2\tt{Q}{H_1}] $ \\ \\ \hline \\

${T_{t3}}={1 \over 4}[6-\tt{G}{Q}-\tt{G}{U}]$ \\ \\

${T_{t2}}={1 \over 4}[-1+4\s{H_2}-2\tt{H_2}{W}-\tt{Q}{W}+4\ttt{H_2}{Q}{W}]$
\\  \\
${T_{t1}}={1 \over 52}[15-18\tt{H_2}{B}+36\s{H_2}-\tt{B}{Q}-12\ttt{B}{Q}{H_2}
-16\tt{B}{U}+48\ttt{B}{U}{H_2}] $ \\ \\

${T_{tt}}={1 \over 12}[6+3\s{H_2}+2\tt{H_2}{Q}+\tt{H_2}{U}] $ \\ \\

${T_{tb}}={1 \over 2}[\s{H_1}+\tt{H_1}{D} ]  $ \\  \\ \hline
\end{tabular}
\end{center}
%%%%%%%%%%%
\begin{center}
\begin{tabular}{|l|}\hline
\multicolumn{1}{|c|}{\bf TABLE II} \\ \hline \\
${Z_{\tau 1}}={3 \over 40} [ 11 +
10\t{B}-8\t{E}-4\tt{B}{E}+8\ttt{B}{E}{H_1}+ $  \\ \\
\hspace{5cm}  $  2\s{H_1}
 -8\st{H_1}{E}-2\t{L}-\tt{B}{L} -8\tt{E}{L}-4\ttt{B}{H_1}{L}
+4\st{H_1}{L}  ]  $
\\  \\
${Z_{\tau 2}}={1 \over 8} [ -3 +
6\s{H_1}-6\t{L}-12\st{H_1}{L}+6\t{W}-3\tt{L}{W}+12\ttt{W}{H_1}{L}] $
\\ \\
${Z_{\tau \tau}}={1 \over 4} [ -16 +
+6\t{H_1}-\tt{H_1}{E}-3\s{H_1} +4\st{H_1}{E} +4\tt{E}{L}
-2\tt{H_1}{L}+ 8\st{H_1}{L}  ]  $ \\ \\ \hline \\
${Z_{b3}}={2 \over 3} [6 -
2\t{D}+4\t{G}-\tt{D}{G}-2\t{Q}-4\tt{D}{Q}-\tt{G}{Q} $ \\ \\
${Z_{b2}}={1 \over 8} [-3
+6\s{H_1}-6\t{Q}-12\st{H_1}{Q}+6\t{W}-3\tt{Q}{W}+12\ttt{W}{H_1}{Q}] $
\\ \\
${Z_{b1}}={1 \over 120} [-21
+10\t{B}-8\t{D}-4\tt{B}{D}+24\ttt{B}{D}{H_1}+18\s{H_1}-24\st{H_1}{D} $ \\ \\
\hspace{5cm} $-2\t{Q}-\tt{Q}{B}+8\tt{Q}{D}+12\ttt{B}{Q}{H_1}-12\st{H_1}{Q}] $
\\ \\
${Z_{bb}}={1 \over 4} [-24
+6\t{H_1}-\tt{H_1}{D}-3\s{H_1}+4\st{H_1}{D}+12\tt{Q}{D}-2\tt{Q}{H_1}
+8\st{H_1}{Q}]$
\\ \\
${Z_{bt}}={1 \over 4} [2\t{H_2}-\s{H_2}-\tt{U}{H_2} ] $
\\ \\ \hline   \\
${Z_{t3}}={2 \over 3} [6 -
2\t{Q}+4\t{G}-\tt{Q}{G}-2\t{U}-4\tt{U}{Q}-\tt{G}{U} ] $ \\ \\
${Z_{t2}}={1 \over 8} [-3
+6\s{H_2}-6\t{Q}-12\st{H_2}{Q}+6\t{W}-3\tt{Q}{W}+12\ttt{W}{H_2}{Q}] $
\\ \\
${Z_{t1}}={1 \over 120} [15
+34\t{B}+18\s{H_2}-2\t{Q}-\tt{B}{Q}-12\ttt{B}{Q}{H_2}+12\st{H_2}{Q}  $ \\ \\
\hspace{5cm}$-32\t{U}-16\tt{B}{U}+48\ttt{B}{U}{H_2}-48\st{H_2}{U}
-16\tt{U}{Q}  ]  $
\\ \\
${Z_{tb}}={1 \over 4} [2\t{H_1}-\s{H_1}-\tt{D}{H_1} ] $
\\ \\
${Z_{tt}}={1 \over 4} [-24 \st{H_1}{U}
+6\t{H_2}-\tt{H_2}{U}-3\s{H_2}+4\st{H_2}{U}+12\tt{Q}{U}-2\tt{Q}{H_2}
+8\st{H_2}{Q}]$
\\ \\ \hline
\end{tabular}
\end{center}

%{\theta_{{\tilde H}_1}}
%{\theta_{\tilde H}}
%{\theta_{{\tilde A}{\tilde B}}}
%{\theta_{{\tilde A}{\tilde B}{\tilde H}}}
%{\theta_{{\tilde A}{\tilde B}{{\tilde H}_1}}}

\end{document}